\documentclass[10pt, a4paper, conference, compsocconf]{IEEEtran}
\usepackage{cite}
\usepackage[pdftex]{graphicx}
\usepackage{epstopdf}
\usepackage{caption} % subfigure
\usepackage{subcaption} % subfigure
\usepackage{url} % \url{my_url_here}

\usepackage{booktabs}
\usepackage{multirow}
\usepackage[para,online,flushleft]{threeparttable}

\usepackage{tabularx}

\begin{document}
\title{The Internet of Things for Aging and Independent Living: A Modeling and Simulation Study}

\author{\IEEEauthorblockN{David Perez, Suejb Memeti, Sabri Pllana\\}
	\IEEEauthorblockA{Department of Computer Science\\
			Linnaeus University\\
			351 95 V\"{a}xj\"{o}, Sweden\\
			\url{dg222ct@student.lnu.se}, \url{suejb.memeti@lnu.se}, \url{sabri.pllana@lnu.se}}
		}

\maketitle

%===========================================================================
\begin{abstract}

Population aging is affecting many countries, Sweden being one of them, and it may lead to a shortage of caregivers for elderly people in near future. Smart interconnected devices known as the Internet of Things may help elderly to live independently at home and avoid unnecessary hospital stay. For instance, a carephone device enables the elderly to establish a communication link with caregivers and ask for help when it is needed. In this paper, we describe a simulation study of the care giving system in the V\"axj\"o municipality in Sweden. The simulation model can be used to address various issues, such as, determining the lack or excess of resources or long waiting times, and study the system behavior when the number of alarms is increased.

\end{abstract}

%===========================================================================
\section{Introduction}

Providing health and social care to the growing population of elderly people poses one of the major societal challenges for the coming decades. Of Sweden’s 9.8 million inhabitants, 20\% are currently above the age of 65 and this number is expected to rise to 23\% by 2040 \cite{care}. About 5.2\% of the population in Sweden is above the age of 80. In 2014, the total cost of elderly care in Sweden was SEK 109.2 billion (EUR 11.7 billion). It is expected that fast networks and pervasive sensor technology will play an important role in meeting the needs of the growing elderly population.

Internet of Things (IoT) refers to a network of various interactive physical objects (such as, personal devices, medical devices, industrial machines, or household goods) with sensing, acting, processing, and communication capabilities \cite{iot-intro}. These interconnected \emph{things} generate big amounts of data that require extreme-scale parallel computing systems for high-performance processing \cite{Reed2015,viebke15,Abraham2015,sp-hybrid-2008,sandrieser12,brandic06,pbb08}. Transition from the Internet of computers to the Internet of things creates opportunities for new services and applications in society, environment and industry \cite{Cerf2016}. One of the application domains of IoT is independent living \cite{aioti15}. IoT may be used to support elderly in their daily activities with reminding services (such as, time to take the medicament, turn off the cooker, close the window when you leave the apartment, location coordinates of things and persons), monitoring services (for instance, state of the chronic diseases), alarming services (establish a communication link with the nurse or doctor in case of emergency) or social services (help to maintain contacts with other people) \cite{EC:IOT}.  

In this paper we focus on IoT solutions for elderly people living in their own houses that are at the point where they become increasingly uncertain about whether they are able to cope with challenges of their daily life. For a city such as V{\"a}xj{\"o}~\cite{vaxjo} with approximately 60 000 inhabitants, about 100 elderly persons leave the hospital every week and up to 80\% of them require home care, such as, support with rehabilitation by a physiotherapist, medical support by a nurse, monitoring the recovery process and supporting the elderly in their daily tasks by welfare helpers. The aim is to enable this significant group of elderly people to live as far as possible independently at their homes, and avoid unnecessary new hospitalizations in future.

\begin{figure}[bt]
\centering
			\includegraphics[width=\linewidth]{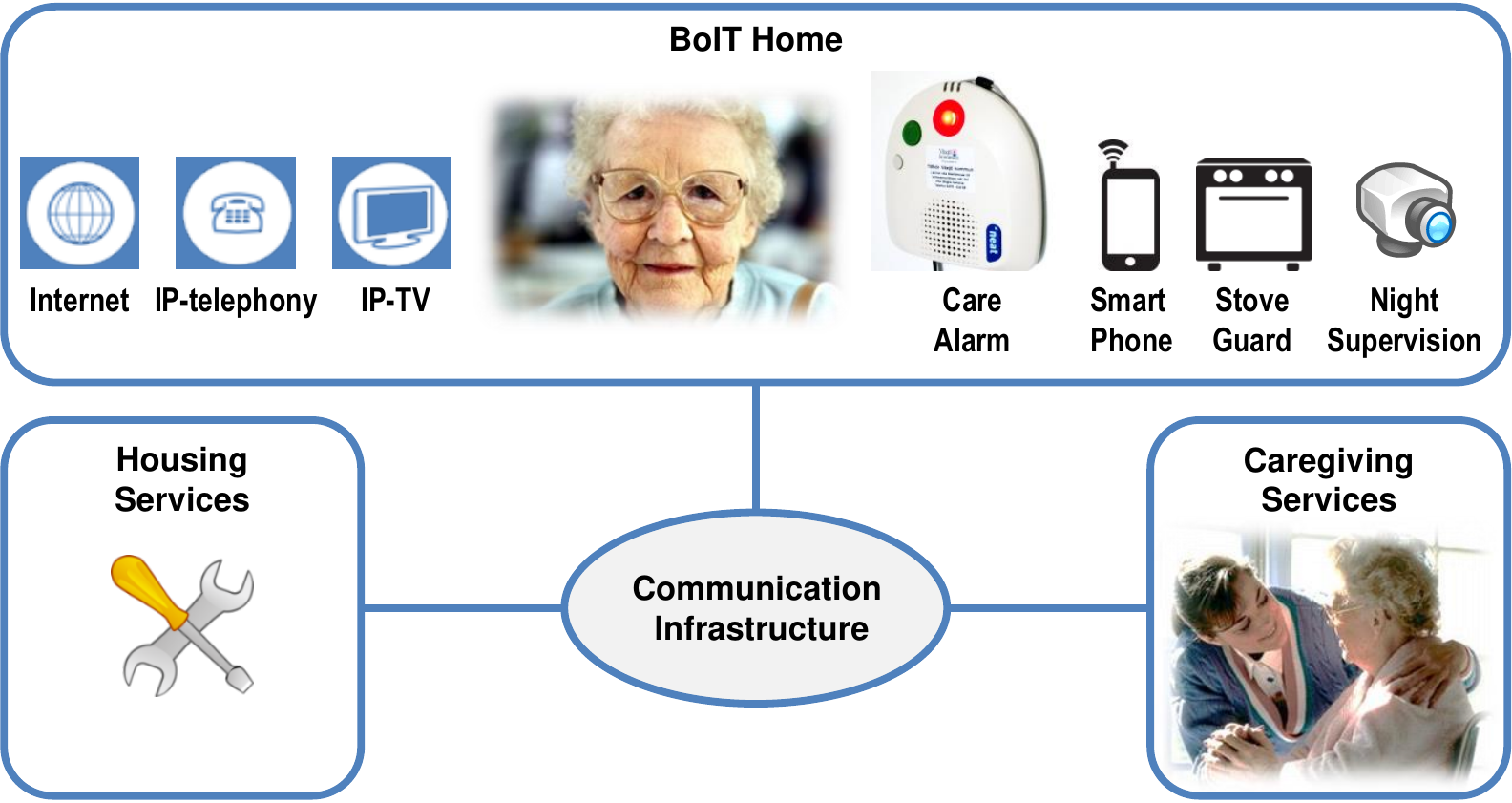}
			\caption{An Internet of Things solution of the BoIT project for independent aging.}
			\label{fig:boit}
\end{figure}

BoIT is a Swedish project of several municipalities including the V\"axj\"o~\cite{vaxjo} municipality that studies the use of IoT solutions for independent living of elderly. Figure~\ref{fig:boit} depicts the BoIT project vision. A BoIT home is equipped with Internet Protocol (IP) telephony, IP TV, care alarm, stove guard, smart phone, night supervision, and various smart household appliances. These devices are interconnected via the Internet and enable the use of remote housing services and care-giving services. 

\begin{figure}[bt]
\centering
  \begin{subfigure}[b]{.5\linewidth}
			\centering
			\includegraphics[width=.9\linewidth]{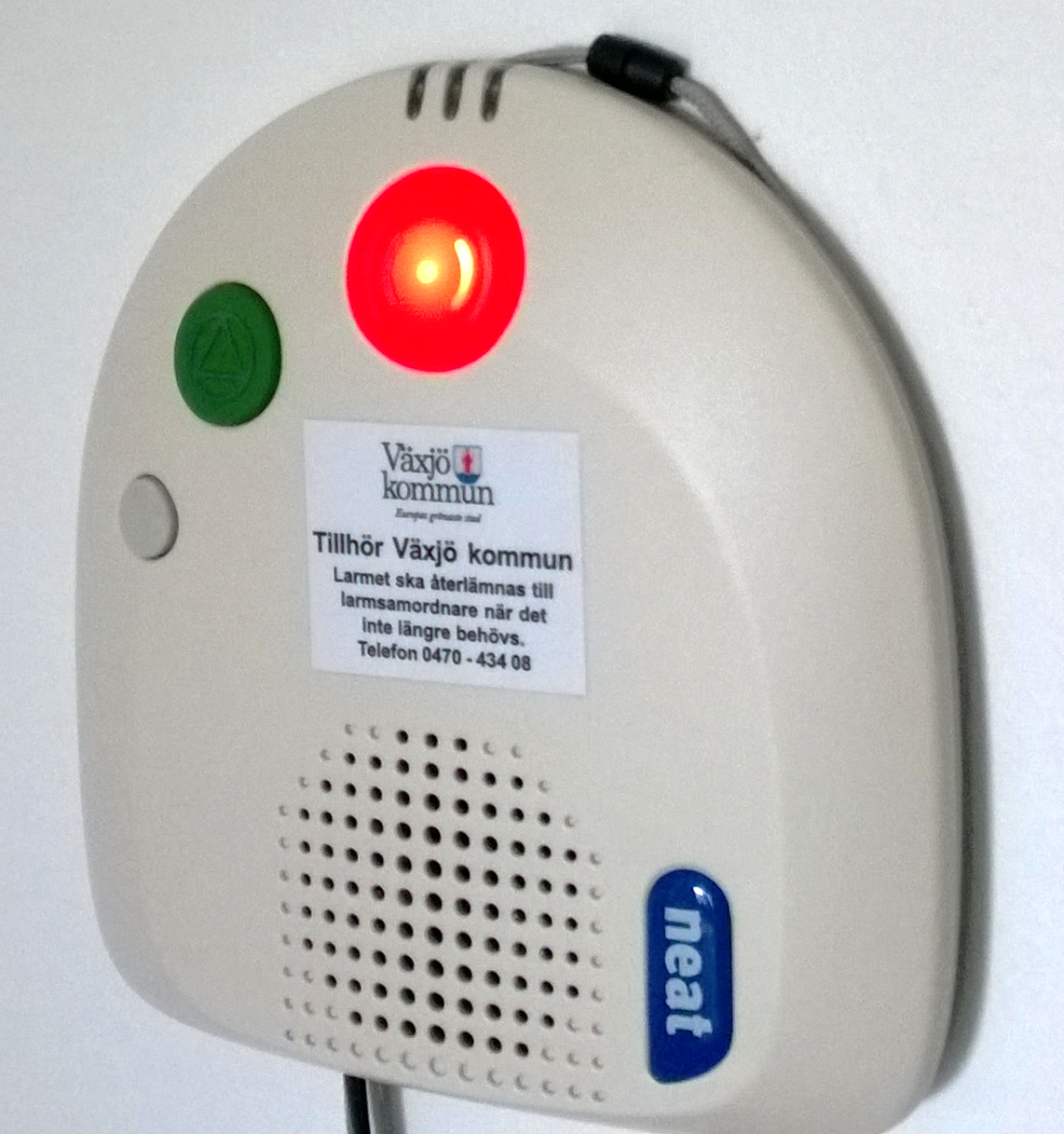}
			\caption{NEO-IP Carephone}
			\label{fig:carephone}
  \end{subfigure}%
  \begin{subfigure}[b]{.5\linewidth}
			\centering
      \includegraphics[width=.5\linewidth]{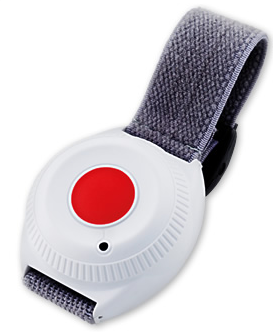}
		  \caption{Fall Detector}
			\label{fig:fall}
  \end{subfigure}
	\caption{NEO-IP Carephone and Fall Detector of NEAT Electronics AB \cite{neat}.}
	\label{fig:iot}
\end{figure}

Our study presented in this paper focuses on the use of care alarm in the context of care-giving services of the V{\"a}xj{\"o}~\cite{vaxjo} municipality. Figure~\ref{fig:iot} depicts the care alarm implementations of NEAT Electronics AB \cite{neat}: NEO-IP Carephone (Figure~\ref{fig:carephone}) and  Fall Detector (Figure~\ref{fig:fall}). NEO-IP Carephone is a care solution for elderly that enables to establish a voice connection via IP with care givers or relatives when the red button is pressed. The Fall Detector is able to detect and report falls automatically. One can send the alarm also manually by pressing the red button of the Fall Detector. 

To study the IoT solution of the V{\"a}xj{\"o} municipality for elderly care we have developed a simulation model using Arena Simulation Software~\cite{SimulationWithArena}. Most of people in V\"axj\"o that use the carephone are between 75 and 80 years old. We received from the V{\"a}xj{\"o} municipality the information about the care giving system (such as, number of carephone user requests per month, allocated personnel to municipality zones, and working schedule) that we used to build and parametrize the simulation model. After the model validation, we used the model to answer what-if questions. For instance, how the system would behave if the arrival rate increases by 5\%, 10\%, or 15\% compared to the actual arrival rate of carephone user requests. Simulation results indicate that a 15\% increase in the arrivals rate would cause unacceptable long waiting times for patients to receive the care.

The rest of the paper is structured as follows. We discuss the related literature in Section~\ref{sec:rw}. Section~\ref{sec:model} describes the development and parametrization of the simulation model. We describe the model validation in Section~\ref{sec:validation}. Section~\ref{sec:results} describes the results of simulation study. Section~\ref{sec:conclusion} concludes the paper and provides ideas for future work. 

%===========================================================================
\section{Related Work}
\label{sec:rw}

Demand for healthcare services is increasing also due to the aging population problem. In the past, scientists have used simulation techniques and software to develop models for analyzing healthcare systems. Information about the importance of simulation studies in healthcare can be found in ~\cite{Lowery1998,Healthcare:Tutorial}.

Most of the related literature addresses similar issues, such as, reducing the waiting time of the patient or reducing costs. For instance, Ballard and Kuhl~\cite{Surgical:Simulation} study a concrete US hospital with the aim of generalizing the model for any hospital. Findlay and Grant~\cite{DES:Outpatient:Healthcare} address the US Army emergency health. Silva and Pinto~\cite{Emergency:Medical:Systems} study the Brazil emergency medical system known as SAMU (Portuguese acronym for Urgency Medical Assistance Service), Weng et al.~\cite{Using:Simulation:and:Data:Envelopment} study the Emergency Department of a Taiwanese Hospital, and Duguay and Chetouane~\cite{Modeling:and:Improving} study the emergency department at Dr. Georges-L. Dumont Hospital in Moncton in Canada. Einzinger et al. ~\cite{GAP-DRG} propose an agent-based simulation model for comparing reimbursement schemes in outpatient care in Austria. 

In contrast to the related work, we use simulation to study an IoT solution for elderly care of the V\"axj\"o municipality in Sweden.

%===========================================================================
\section{Development of the Simulation Model with Arena}
\label{sec:model}

\begin{figure}[ht]
\centering
			\includegraphics[width=\linewidth]{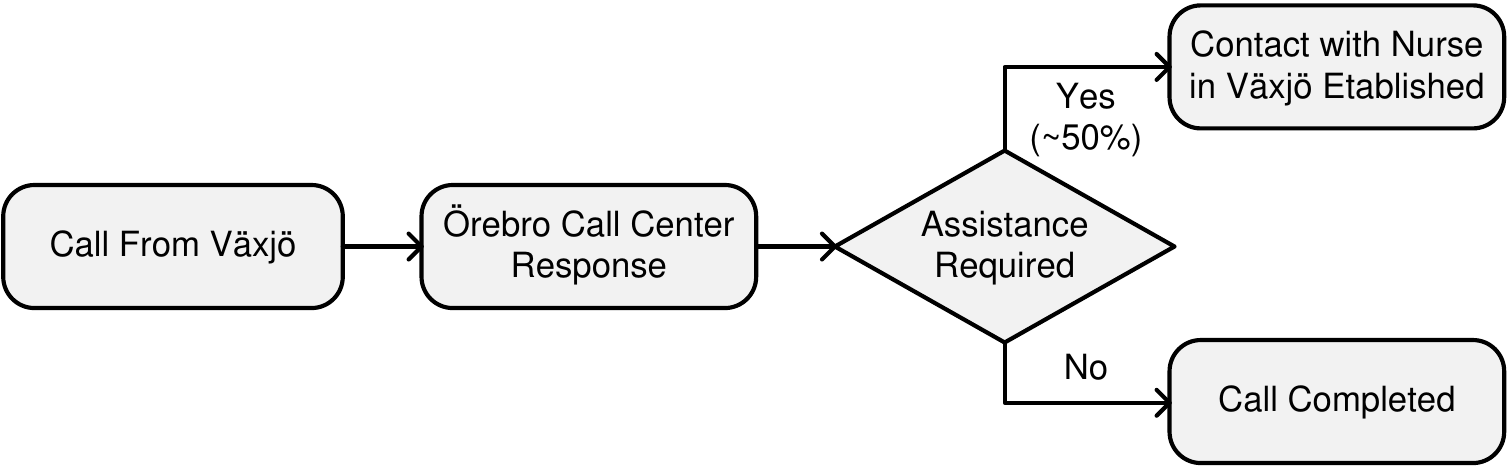}
			\caption{Overview of the Carephone call process.}
			\label{fig:call_process}
\end{figure}

Figure~\ref{fig:call_process} depicts the procedure for handling requests for assistance in V\"axj\"o. When a user presses the red button of the IP carephone, a voice connection with the call center in \"Orebro is established. The operator in the call center will establish the contact with a nurse in the zone of patient in V\"axj\"o if assistance is required. During the day the care assistants are organized in care groups, each group taking care of one of the 18 zones in V\"axj\"o (Table~\ref{table:Care_groups_staff}). During the night each of the 18 zones is associated to one of the four night patrols. A night patrol has six care assistants. The nurse will spend between 4 and 10 minutes depending on the current location to get to the patient's home. Usually the nurse spends from 10 to 60 minutes to assist the patient at his location.

\begin{table}[tb]
\caption{Personnel of the care groups for each zone in V\"axj\"o.}
\begin{center}
\begin{tabular}{ l r r }
\toprule
Zone & Weekday Personnel & Weekend Personnel \\ 
\midrule
~1. Anna Trolle & 19 & 11\\ 
~2. Dalbo & 12& 7\\
~3. Teleborg  &  13& 9\\ 
~4. Rottne & 9& 9\\ 
~5. Centrum & 16& 8\\ 
~6. S{\"o}der & 10& 7\\ 
~7. Lammhult & 15& 6\\ 
~8. Lassaskog & 12& 7\\ 
~9. {\"O}jaby & 10& 6\\ 
10. Sandsbro & 10& 6\\ 
11. Hovslund & 7& 6\\ 
12. Borgm\"{a}staren & 9& 5\\ 
13. {\"O}ster & 10& 6\\ 
14. {\"O}ster {\AA}ryd & 5& 3\\ 
15. Kinnevald & 9& 5\\ 
16. Gemla & 5& 4\\ 
17. Kvarng{\aa}rden & 10& 5\\ 
18. Sj{\"o}liden & 9& 4\\ 
\bottomrule
\end{tabular}
\end{center}
\label{table:Care_groups_staff}
\end{table}

\begin{table}[tb]
\caption{V\"axj\"o zones assigned to each Night Patrol.}
\begin{center}
\begin{tabularx}{\linewidth}{ l X }
\toprule
Night Patrol & Zones Assigned \\ 
\midrule
1. North & Sj{\"o}liden, Rottne and Lammhult \\ 
2. East & Sandsbro, Hovslund, {\"O}ster, {\"O}ster {\AA}ryd and Anna Trolle \\ 
3. South  & Dalbo, S{\"o}der, Kvarng{\aa}rden and Teleborg  \\ 
4. West & Centrum, Lassaskog, Borgm{\"a}staren, {\"O}jaby, Gemla and Kinnevald \\ 
\bottomrule
\end{tabularx}
\end{center}
\label{table:Night_Patrols_Dist}
\end{table}

With respect to the choice of simulation environment we have studied the ARGESIM Benchmarks \cite{argesim,pllana-c6}. Currently there are about 20 ARGESIM benchmarks that are used to compare simulation languages/environments considering various features, such as, modeling techniques, event handling, distribution fitting,  output analysis, animation, or complex logic approaches. For implementation of the simulation study we have selected Arena \cite{SimulationWithArena}, which is a discrete-event simulation environment offered by Rockwell Automation. Arena concepts include: entities, attributes, variables, resources, queues, statistical accumulators, submodels, simulation clock. In the following we highlight some aspects of our simulation model of the care-giving system in V{\"a}xj{\"o}.

\emph{Entities.} In our simulation model the \textit{Patient} is modeled as Arena entity. During the simulation this entity will seize and release resources that represent in our model the people (such as, nurses or call center operators) in the care-giving system.

\emph{Resources.} Resources in our simulation model are nurses groups and the call center staff. Nurses are grouped by their schedule (day or the evening, weekday or the weekend) and by the zone they work. Available care resources grouped according to zones in V\"axj\"o are listed in Table \ref{table:Care_groups_staff}.

\emph{Attributes.} The attribute \textit{nZone} in the simulation model defines the zone that the patient belongs to. This attribute is integer number in the range from 1 to 18, each number corresponding to one of the 18 zones in V\"axj\"o. 

\begin{figure}[b]
\center
\includegraphics[width=\linewidth]{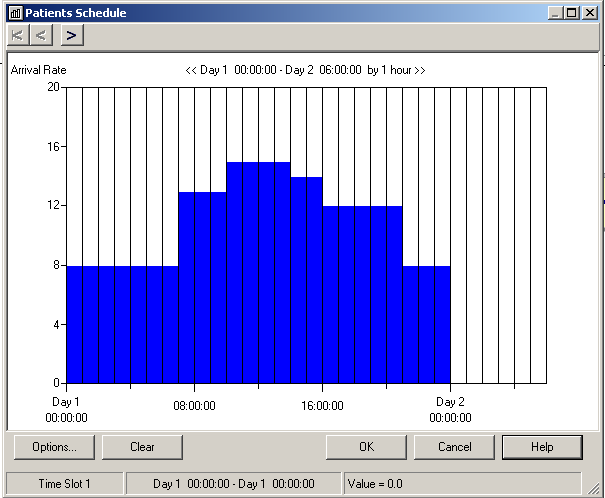}
\caption{The schedule for patient call arrival.}
\label{fig:Schedule}
\end{figure}

\emph{Schedules.} Schedules in Arena enable to define entities arrival using a graphical interface. In our simulation model the schedule is defined by statistics about the patient calls made in V{\"a}xj{\"o} during the November 2013. Figure~\ref{fig:Schedule} depicts the Arena schedule for patient call arrival in our simulation model. 

\begin{figure*}[t]
\center
\includegraphics[width=0.6\linewidth]{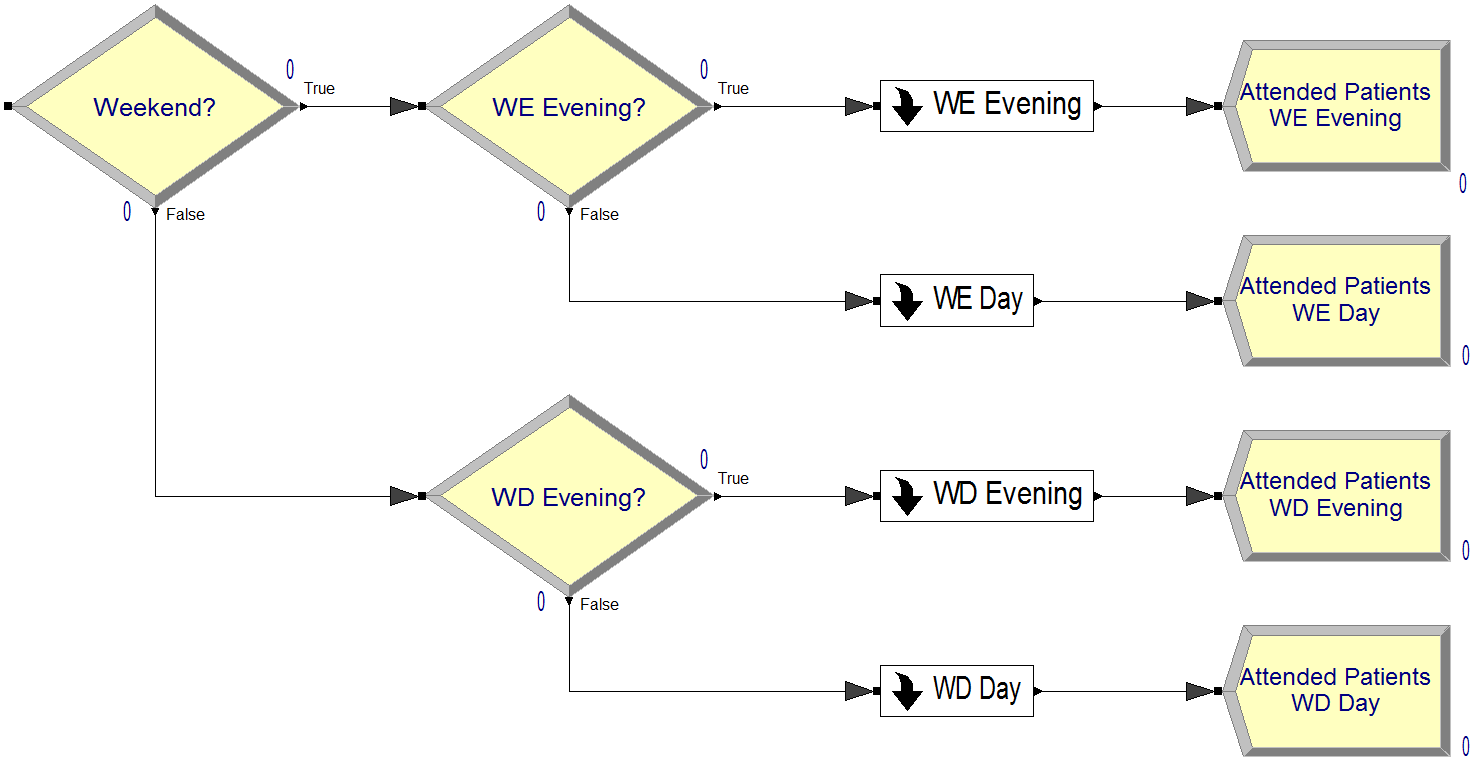}
\caption{Arena simulation submodel for the assistance procedure of care-giving system in V{\"a}xj{\"o}. WD stands for weekday (Monday to Friday), WE stands for weekend.}
\label{fig:Assistance_Part}
\end{figure*}

\begin{figure}[bt]
\center
\includegraphics[width=\linewidth]{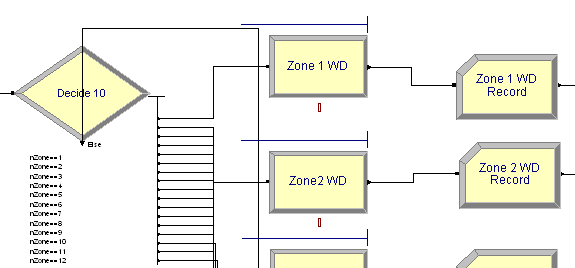}
\caption{Part of Arena submodel for handling patients according to city zones during weekdays (Monday to Friday).}
\label{fig:During_Day_Submodel}
\end{figure}

Figure~\ref{fig:Assistance_Part} depicts the submodel in Arena that models the assistance procedure of care-giving system in V{\"a}xj{\"o}. The first \emph{Decide} module checks if the entity arrives during weekdays (Monday to Friday) or weekends. WD stands for weekday, WE stands for weekend. We use the Arena Variable \textit{TNOW} to store the time in hours during the simulation. The total number of hours of the week is 168. If $MOD(TNOW,168)>120$ then it is weekend, otherwise it is a weekday. The next \emph{Decide} module is used to check if the call has been made during the day or during the evening. If $MOD(TNOW,24)$ is greater than 21 and less than 7 then the call arrived during the evening. If the call arrived in the evening then the \emph{Evening} submodel handles the simulation, otherwise the \emph{Day} submodel (Figure \ref{fig:During_Day_Submodel}) is selected for simulation. During the day are available 18 care groups corresponding to 18 zones of V{\"a}xj{\"o}, whereas in the evening are available four night patrols (Table \ref{table:Night_Patrols_Dist}). For instance, Night Patrol 1 is responsible for patients from zones 4, 7, and 18. We determine the responsibility of Night Patrol 1 using the following expression $(nZone == 4) || (nZone == 7) || (nZone == 18)$.

%===========================================================================
\section{Validation}
\label{sec:validation}

To verify and validate the model, first the care giving system in V{\"a}xj{\"o was examined. We had a series of meetings with an assistance nurse for the carephone users and the system administrator to collect as much information as possible about how the system works (such as, number of calls, number of nurses, number of staff assisting the calls). We kept contact with the system administrator while the simulation model was being built to ensure that the model conforms to the real care giving system. 

The model implementation can be checked by the simulation software itself, and it is always checked before any simulation takes place. If there are model implementation errors the simulation will not run. 

After running initial test simulations and checked that the results made sense, the same validation system used in ~\cite{DES:Outpatient:Healthcare, Emergency:Medical:Systems, Modeling:and:Improving} was used to validate our model. That is, running 100 replicas of a whole month simulation, and building a confidence interval with a confidence level of 95\% with all the data collected from that simulation. After that, we checked that the real values are inside confidence intervals.

We can observe in Table~\ref{table:Confidence}, where the real values and the upper and lower bounds of the confidence interval are shown, the value from each time interval of the real data is within the confidence interval built from the simulation results. 

\begin{table}[tb]
\caption{Patients per month. Comparison of real data with the simulated data; 95\% confidence level.}
\label{table:Confidence}
\begin{center}
\begin{tabular}{ r  r  r  r }
\toprule
 Period & Real data & Upper Bound & Lower Bound \\ 
\midrule
07:00 to 10:00 & 1193 & 1200 & 1187 \\ 
10:00 to 14:00 & 1776 & 1780 & 1764 \\ 
14:00 to 16:00  & 860 & 862 & 851 \\ 
16:00 to 21:00 & 1750 & 1769 & 1750 \\ 
21:00 to 07:00 & 2284 & 2287 & 2269 \\ 
\midrule
Total & 7863 & 7881 & 7841 \\ 
\bottomrule
\end{tabular}
\end{center}
\end{table}

%===========================================================================
\section{Results}
\label{sec:results}

In this section we discuss and analyze the obtained results with respect to the behavior of entities (\emph{patients}), resources (\emph{care giving personnel}) and queues. 

%---------------------------------------------------------------------------
\subsection{Results and Analysis of Patient Waiting Times}
\label{sec:results-entities}

The patient waiting time is an important aspect of an emergency system. We have adopted a method described in Ballard and Kuhl~\cite{Surgical:Simulation} and applied to our model, which means increasing the number of arrivals by 5\% at a time until the waiting time for patients is not suitable for the emergency care system.

The maximum waiting time that is considered suitable for the patient is 25 minutes. The waiting time includes:
\begin{itemize}
	\item Call Center Time - that is the time spent for the patient contacting the call center, and the time spent by the call center operator contacting the nurse.
	\item Transfer Time - that is the time spent by the nurse going from one location to another. In our case this is the time spent by the nurse going to the patient's location.
\end{itemize}

Table \ref{tab:waiting-times} lists the waiting time until the patient receives care services for various arrival rates. We may observe that the \emph{Call Center Time} increases with the increase of arrival rate. The \emph{Transfer Time} is approximately 10 minutes (please note that we consider the worst case scenarios) in every simulated experiment. Results show that the total waiting time with real arrivals data is 15 minutes, which is still acceptable for our system. If the number of arrivals increases by 5\% and 10\% the response time will be 16.8 and 19.2 minutes. Further increase of the arrivals to 15\% results with unacceptable waiting times (above 25 minutes) for an emergency care giving system. 
	
\begin{table}[tb]
	\caption{The waiting times (in minutes) for the patient to receive care services. We determine the maximum capacity of the entities that is suitable for the care giving system increasing the number of arrivals by 5\% at a time.}
	\centering
	\begin{tabular}{rcccc}
		\toprule
		& Real Arrivals & +5\%  & +10\% & +15\%  \\ 
		\midrule
		Call Center Time & 5.4 & 7.2 & 9.6 & 18 \\ 
		Transfer Time & 9.6 & 9.6 & 9.6 & 9.6 \\ 
		\midrule
		Total Waiting Time  & 15 & 16.8 & 19.2 & 27.6 \\ 
		\bottomrule
	\end{tabular}
	\label{tab:waiting-times}
\end{table}

Figure \ref{fig:assistance-time} shows the impact of increasing the number of arrivals on the Total Waiting Time. We may observe that once the available resources can not cope with the number of requests, the waiting time will increase with a steep slope and exceeds the limit of 25 minutes. 

\begin{figure}[htb]
	\centering
	\includegraphics[width=\linewidth]{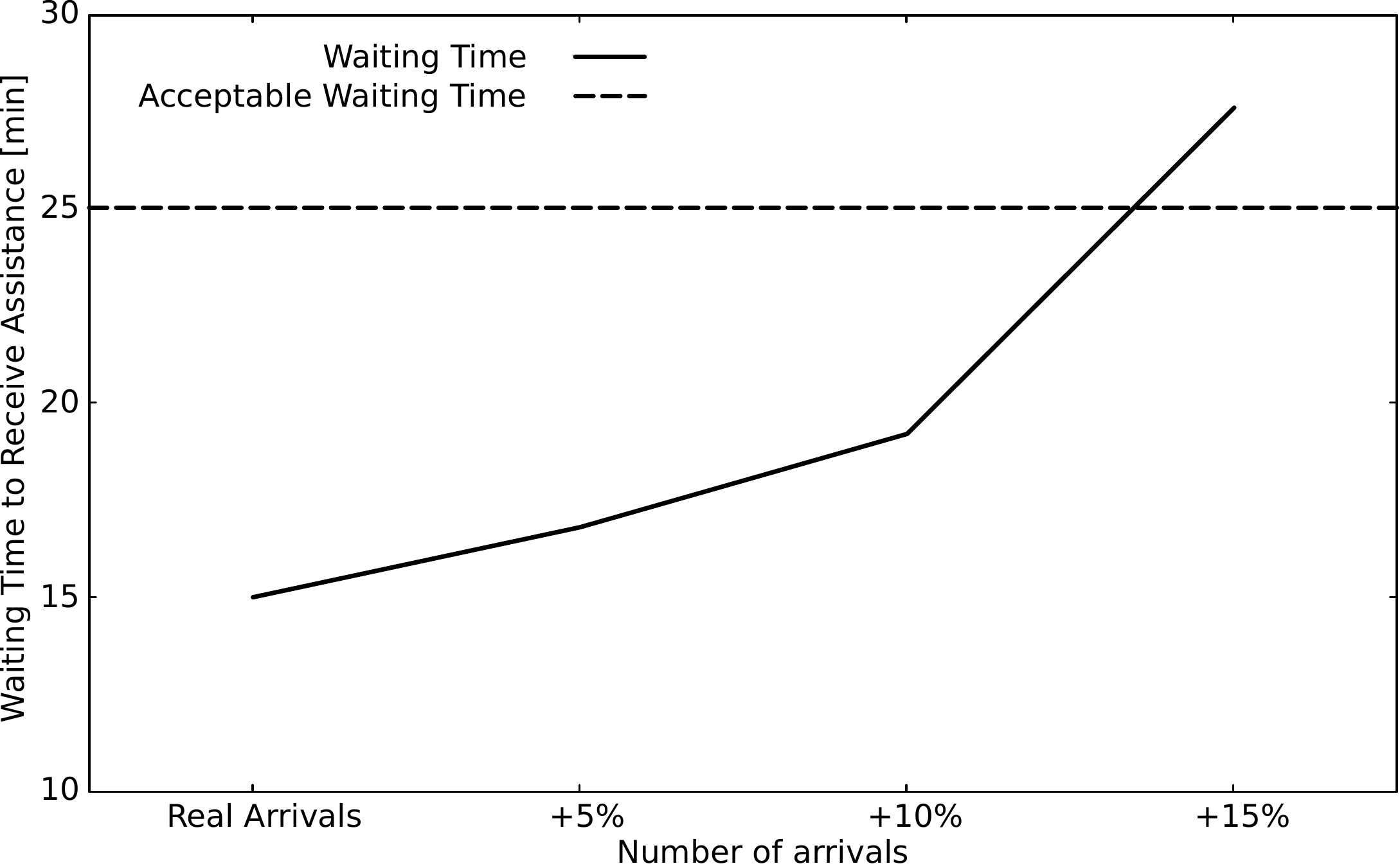}
	\caption{The impact of increase of number of arrivals on the total waiting time for the patient to receive assistance.}
	\label{fig:assistance-time}
\end{figure}

%---------------------------------------------------------------------------
\subsection{Analysis of Resource Utilization}
\label{sec:results-resources}

The experimental results (see Table \ref{tab:night_usage}) indicate that the Night Patrol resources are highly utilized and will most likely collapse in case arrivals increase. Even though the number of the arrivals decreases almost by 50\% during the night, the number of the personnel is reduced as well. Furthermore, the night patrols operate on multiple (3 to 6) different zones. 

\begin{table}[tb]
	\caption{Resource utilization of night patrols expressed in percentage.}
	\centering
	\begin{tabular}{lr}
		\toprule
		Night Patrol & Usage Percentage \\ \midrule
		East   & 100\% \\ 
		West   & 100\% \\ 
		North  & 83\% \\
		South  & 100\%\\ \bottomrule
	\end{tabular}
	\label{tab:night_usage}
\end{table}

Table \ref{tab:res-util-wd-we} lists resource utilization for various zones expressed in percentage \% during weekdays (Monday to Friday) and weekends. We may observe that most of the resources are less than 80\% utilized. Due to the decrease of the personnel during weekends, the utilization is higher compared to weekdays. Some of the zones that might collapse in case of arrivals increase are \"{O}ster \AA{}ryd, Hovslund and Borgm\"{a}staren that are 100\%, 83.3\% and 80\% utilized during weekends. 

\begin{table}[tb]
	\caption{Resource utilization expressed in percentage for each of the zones during weekdays (Monday to Friday) and weekends.}
	\centering
	\label{tab:res-util-wd-we}
	\begin{threeparttable}
		\begin{tabular}{@{}lrr@{}}
			\toprule
			 \multirow{2}{*}{} & \multicolumn{2}{l}{Utilization percentage}\\ \cmidrule(l){2-3} 
										              & weekdays   & weekends  \\ \midrule
			Anna Trolle\tnote{2}		    & 26\%        & 45\%     \\
			Dalbo\tnote{3}        		  & 50\%        & 71\%     \\
			Teleborg\tnote{3}     		  & 38\%        & 56\%     \\
			Rottne\tnote{1}       		  & 44\%        & 44\%     \\
			Centrum\tnote{4}      		  & 25\%        & 50\%     \\
			S\"{o}der\tnote{3}        	& 50\%        & 71\%     \\
			Lammhult\tnote{1}     		  & 27\%        & 67\%     \\
			Lassaskog\tnote{4}     		  & 33\%        & 57\%     \\
			\"{O}jaby\tnote{4}          & 40\%        & 67\%     \\
			Sandsbro\tnote{2}     		  & 40\%        & 50\%     \\
			Hovslund\tnote{2}     		  & 57\%        & 83\%     \\
			Borgm\"{a}staren\tnote{4} 	& 30\%        & 80\%     \\
			\"{O}ster\tnote{2}        	& 40\%        & 67\%     \\
			\"{O}ster \AA{}ryd\tnote{2} & 60\%        & 100\%    \\
			Kinnevald\tnote{4}    		  & 44\%        & 60\%     \\
			Gemla\tnote{4}        		  & 80\%        & 75\%     \\
			Kvarng\aa{}rden\tnote{3}    & 30\%        & 60\%     \\
			Sj\"{o}liden\tnote{1}     	& 33\%        & 75\%     \\
			\bottomrule
		\end{tabular}
		\begin{tablenotes}
			\item[1] North Night Patrol; 
			\item[2] East Night Patrol;
			\item[3] South Night Patrol;
			\item[4] West Night Patrol;
		\end{tablenotes}
	\end{threeparttable}
\end{table}

%---------------------------------------------------------------------------
\subsection{Analysis of Queue Waiting Times}
\label{sec:results-queues}

In an emergency system, it is important that queue waiting times are not large. Since typical assistance times range from 10 to 60 minutes, a single patient in the queue may result with a critical situation. In the system under study, during the night patients may have to wait in queues and during the day the queue lengths are zero. The maximal observed queue waiting time for the \emph{East Night Patrol} during weekdays is 5.1 minutes, whereas during weekends the waiting time slightly increases to 5.5 minutes due to decrease of the available care personnel. 

Considering the required time to contact the call center, time to contact the nurse, and the time spent by the nurse going to the patient's home, an additional 5 minutes waiting time for resources in the queue can be critical in emergency cases. 

Results indicate that the amount of resources in care giving system is appropriate for the current rate of arrivals. Redistribution of the staff, moving some of the personnel from the weekdays care groups to other groups (for instance, night patrols) or some of the zones where the utilization is higher, would be a good way of preventing long queues in case the arrivals increased. On the other hand, as the utilization is not generally high, and only one nurse at a time is needed to take care of a patient, increasing the number of nurses with the current rate of arrivals would not increase the quality of the service.

%===========================================================================
\section{Conclusion and Future Work}
\label{sec:conclusion}

In this paper we have described an IoT solution for elderly care of the V\"axj\"o municipality in Sweden. To study the system we developed a simulation model using Arena Simulation Software. We validated the simulation model by running a series of simulations, and checking whether the real data is within the confidence intervals. 

We performed several simulation experiments to observe different aspects of the system.

\begin{enumerate}
\item Studied the behavior of the system when the arrival rate increases. We observed that a 15\% increase in the arrivals rate would cause unacceptable long waiting times for patients to receive the care. For the V{\"a}xj{\"o} elderly care giving system is important to know that additional personnel would be needed when such an arrival rate increase occurs.

\item Studied the resource utilization. Currently, there are no significant bottlenecks in the system, and the available resources are sufficient to handle the current arrival rate of patient requests. 

\item Studied the queue behavior. Since it is an emergency care giving system, it is desired that queue lengths are zero. Simulation revealed that in the worst case scenario, some resources do have a short queue (which is only 1 patient long for all the cases), and even being as short as it is, it generates waiting times of about 5 minutes. Depending on disease, those additional 5 minutes may be critical in an emergency situation 

\end{enumerate}

Based on simulation results we conclude that the care giving system under study is efficient with respect to the current arrival request rate in most circumstances. Exceptions include the night patrols and a municipality zone where the utilization is near 100\% in the worst case scenario. According to the simulation results, a minor staff redistribution could be reasonable. 

Although the data collected from the real-world system was enough to study the system operation and resource usage, the availability of more accurate data about the individual transportation times and times spent taking care of the patient could further improve the simulation study in future.

The simulation model could be extended in future. For instance, in critical cases where the patient needs assistance beyond what a nurse can provide at home, an ambulance would be called and the patient might be hospitalized. The model could be extended with with a number of ambulances and hospital rooms as resources.


\end{document}